\documentclass[conference]{IEEEtran}
\ifCLASSINFOpdf
\else
\fi

\usepackage{amsmath}
\usepackage{algorithm}
\usepackage{algorithmic}
\usepackage{multirow}
\usepackage{array,graphicx}
\usepackage{pifont}
\newcommand*\rot{\rotatebox{90}}
\usepackage[english]{babel}
\usepackage[autostyle]{csquotes}
\usepackage{url}

\begin{document}
%
\title{Predicting Small Group Accretion in Social Networks: A topology based incremental approach}

\author{\IEEEauthorblockN{Ankit Sharma, Rui Kuang and Jaideep Srivastava}
\IEEEauthorblockA{Department of Computer Science and Engineering\\
University of Minnesota, Minneapolis, MN 55455\\ \{ankit,kuang,srivasta\}@cs.umn.edu}
\and
\IEEEauthorblockN{Xiaodong Feng}
\IEEEauthorblockA{University of Electronic Science \\ and Technology of China, China\\
 fxdong88@gmail.com}
\and
\IEEEauthorblockN{Kartik Singhal}
\IEEEauthorblockA{LNM Institute of Information \\ Technology, Jaipur, India \\ singh559@umn.edu}}



\maketitle

\begin{abstract}
Small Group evolution has been of central importance in social sciences and also in the industry for understanding dynamics of team formation. While most of research works studying groups deal at a macro level with evolution of arbitrary size communities, in this paper we restrict ourselves to studying evolution of small group (size $\leq20$) which is governed by contrasting sociological phenomenon. Given a previous history of group collaboration between a set of actors, we address the problem of predicting likely future group collaborations. Unfortunately, predicting groups requires choosing from $n \choose r$ possibilities (where $r$ is group size and $n$ is total number of actors), which becomes computationally intractable as group size increases. However, our statistical analysis of a real world dataset has shown that two processes: an external actor joining an existing group (\textit{incremental accretion (IA)}) or collaborating with a subset of actors of an exiting group (\textit{subgroup accretion (SA)}), are largely responsible for future group formation. This helps to drastically reduce the $n\choose r$ possibilities. We therefore, model the attachment of a group for different actors outside this group. In this paper, we have built three topology based prediction models to study these phenomena. The performance of these models is evaluated using extensive experiments over DBLP dataset. Our prediction results shows that the proposed models are significantly useful for future group predictions both for IA and SA.
\end{abstract}
\vspace{-0.3em}
\begin{IEEEkeywords}
Social Networks, Higher Order Link Prediction, Group Evolution, Hypergraphs, Hypergraph Evolution
\end{IEEEkeywords}
\IEEEpeerreviewmaketitle

\vspace{-1.3em}
\section{Introduction}

Study of small groups has been an important endeavor in a large number of disciplines like psychology, sociology, communication and information science for past 50 years \cite{poole2004theories}. Advent of globalization has lead to changing nature of groups or teams in industry leading to increasing interest in studying their dynamics \cite{cheney2010organizational}. Large part of the research in the field of Organization Science is dedicated to study effective ways to build teams by combining employee expertise \cite{boh2007expertise}. With the rising number of large interdisciplinary scientific teams \cite{wagner2011approaches}, understanding drivers affecting their success is of key importance for science funding agencies while selecting team of scientists \cite{lungeanu2014understanding}. Other real life applications include building emergency response teams for natural disasters management, automation of team selection for military operations and self-organizing open-software teams \cite{hahn2008emergence}. While such studies are important, the ever increasing availability of online \enquote{group} interaction data for example, social networking sites like Facebook or Twitter, group communication tools like Skype, Google Hangout, Google Docs, Massive Online multi-player games (MMOGs) such as World of Warcraft, etc., makes such studies even more realistic. In scientific research, such dataset have been used to study group dynamics for benefit of both industry and academia \cite{contractor2013some}.


\begin{figure}[h!]
\centering
\includegraphics[width=50mm]{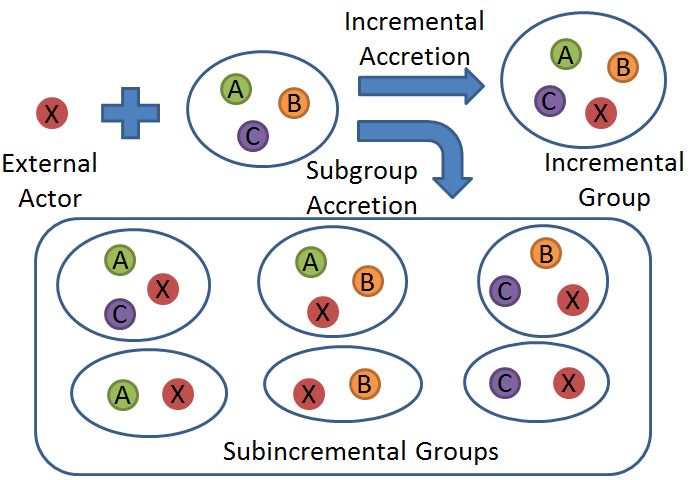}
\caption{Example illustrating \textit{incremental accretion} and \textit{subgroup accretion} processes. Set \{A,B,C\} is a group and X is an external actor.}\label{fig:basicprocessesfig}
\label{overflow}
\vspace{-0.9em}
\end{figure}

\begin{figure}[h!]
\centering
\includegraphics[width=60mm,height=2.4cm]{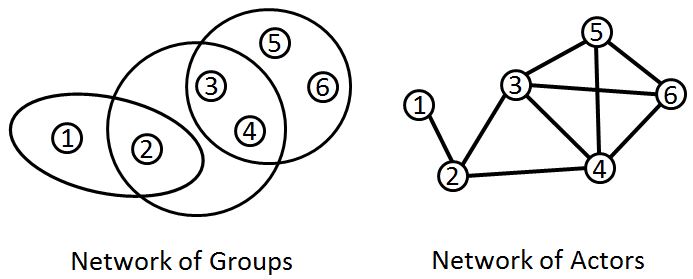}
\caption{Example illustrating \textit{network of groups} (left) and the corresponding \textit{network of actors} (right) where \{1,2,3,4,5,6\} are the actors}\label{fig:NOGNOA}
\label{overflow}
\vspace{-0.9em}
\end{figure}

Several studies in the past have analyzed the usefulness of various features used for membership prediction in communities \cite{patil2013stability}\cite{patil2012destruction}\cite{sharara2012stability}\cite{kang2013loyalty}. Also, some works have focused on group membership dynamics by simulating how individual join or leave a group \cite{johnson2009human}\cite{ahmad2011guild}\cite{alvari2014predicting}, while such studies have not developed prediction models. Objective of our work is to focus on the task of actual future group prediction in contrast to feature analysis or simulation based studies. Moreover, most past works on group evolution in social networks primarily deal with evolution of arbitrary size communities or groups \cite{chen2008player}\cite{patil2013attrition}\cite{sharara2009dynamics}. These sizes are usually large and the boundaries of community depend on the definition of membership employed \cite{spiliopoulou2011evolution}.  In this work we are interested in well defined small groups (size $\leq20$) like research collaborations or teams \cite{beebe2009communication} also called as \textit{Bona fide groups} \cite{putnam2012bona} in sociology. Also these small groups are self assembling \cite{contractor2013some} where members leave or join groups autonomously and the motivation or theories for group formation is much different from the large communities \cite{poole2004theories}. Moreover, these groups are connected by the social network of actors, resulting in group ties or \textit{network of groups} \cite{monge2003theories} (see Figure~\ref{fig:NOGNOA}), which plays central role in group formation process. We focus on group accretion which is a subset of the group evolution. Group accretion is the process of size increment in groups by addition of more members. More specifically we define two subproblems: \textit{incremental accretion} and \textit{subgroup accretion}. In the first problem, given a group of size=$x$, we predict the likelihood of one more member being absorbed in it, to yield a size=$(x+1)$ incremental group (Figure~\ref{fig:basicprocessesfig}). The subgroup accretion is the problem of incremental accretion on all the $(2^x -2)$ subgroups of a given group to yield prediction scores of $(2^x-2)$ new incremental groups (Figure~\ref{fig:basicprocessesfig}). Intuition behind choice of these problems is that, given a past history of group collaborations, a large percentage of groups in future are formed through these two processes. Our aim finally is to build models that predict future groups that are likely to be formed using these two mechanisms. This work therefore, is an initial step towards a more general higher order group prediction problem.           

In this paper we assume that the groups are not isolated but rather interact with each other through \textit{network of groups} and group members make individual decisions to collaborate or not (\textit{self assembly}). We therefore, model the attachment of a group (\textit{group}) to an actor (\textit{node}) outside it. Topology based dyadic link prediction (DLP) methods have proven successful in capturing \textit{node to node} attachment. Guided by both, DLP methods and sociology theories of small groups we have proposed three different methods. First, method is an extension of the popular path enumerating Katz method \cite{katz1953new}. Second, is a network alignment based supervised method. With a philosophy similar to Katz, it captures the inter-group communication cycles, which are theoretically hypothesized more appropriate, rather than paths. Lastly, we also propose a label propagation based method where the random walks are guided by the network of groups (a hypergraph \cite{berge1973graphs}). Based on the extensive set of experiments it turns out our models are able to effectively predict future groups formed by both IA and SA processes.
\vspace{-0.5em}

Now we summarize the key points of this research paper:

\begin{itemize}
  \item We highlight an important observation that a large chunk of future groups are formed by two different accretion processes from the past groups.
  \item We have focused on the evolution of small groups and defined new problems for small group accretion. 
  \item We have developed three topology based methods, guided by social theories, to address these problems in a novel manner.
  \item We have therefore, presented an overall new incremental approach to address the less addressed and intriguing problem of higher order link prediction. 
\end{itemize}
\vspace{-0.5em}

\section{Related Works}

\textbf{Social Group Evolution}
There is a vast body of literature regarding community evolution \cite{spiliopoulou2011evolution} and more general network evolution \cite{aggarwal2014evolutionary} in social networks. Definition of community in all these works varies drastically producing from small groups to size as large as hundreds or thousands. Therefore, they take a macroscopic view while answering the questions of how to detect communities and how they evolve over time. Though, recently there have been some works which zoom in and try to understand the evolution from the perspective of individual actors and their relationship with other actor group. 
\vspace{-0.5em}

Among them the first category of works focus on building models which can simulate the group formation tasks like leaving, joining or switching between groups \cite{johnson2009human}\cite{ahmad2011guild}\cite{alvari2014predicting}. Alvari et. al. \cite{alvari2011detecting} provide a game theoretic community/group detection model where the actors in the social network are rational agents performing these tasks while maximizing their utility. Same authors extend their work for evolutionary setting \cite{alvaricommunity} and apply to MMOGs \cite{alvari2014predicting}. MMOG guild formation is also studied as stochastic processes in both network \cite{ahmad2011guild} and network-less settings \cite{johnson2009human}. Our work, rather than simulation, focuses on predicting the exact groups that might occur in future.
\vspace{-0.5em}

Second category deals with analysis of various characteristics of groups (like diversity, cohesion, stability, type, etc.) and their correlation with different factors (size, member properties, etc.) \cite{chung2014}\cite{chen2008player}.
\vspace{-0.5em}

Third category of papers are devoted to extracting features of different kinds like network, actor, group or communication content and pose the problem of group membership prediction as a classification task. For instance, Patil et al. have done feature extraction for group attrition \cite{patil2013attrition}, group stability \cite{patil2013stability} as well as group destruction \cite{patil2012destruction}. It differs from our work as they focus on understanding the importance of various features rather than prediction of groups. In a series of papers from Sharara et al. have stressed on the idea of loyalty (affinity) of actor towards different groups and its longitudinal changes \cite{sharara2009dynamics}. This idea has been extended for deducing important actors by using group semantics (like diversity) \cite{sharara2012stability} and applied to analysis of guilds in MOMGs \cite{kang2013loyalty}. Both, Patil et al. and Sharara et al. model actor's attachment or loyalty for different groups whereas we model the opposite: tendency of a group to absorb different actors. Moreover, both of them primarily deal with large communities like conferences in DBLP data. We are specifically focused on predicting small cohesive groups or teams (like a small group of researcher working on a publication in DBLP) where different social phenomenon are at play. We rather treat the problem as extension of DLP to higher order (group) prediction.   
\vspace{-0.5em}

\textbf{Social Sciences} Naturally occurring small groups are called \textit{Bona Fide Groups} \cite{putnam2012bona}, which are the focus of this paper. A good reference for small group theories can be found in Poole et al. \cite{poole2004theories}. Plethora of research deals with application of small group dynamics for understanding teams \cite{boh2007expertise}. Network perspective of teams has been a new development in the past decade. Various studies like Oh et. al. \cite{oh2004group}\cite{lungeanu2014understanding} stress the importance of network structure in determining team performance. More recent research has focused on self-assembling teams in which members autonomously leave or join teams \cite{contractor2013some}\cite{lungeanu2014understanding}. 
\vspace{-0.5em}
    
\textbf{Link Prediction} Due to space constraint we point out some key surveys and papers for DLP. An overview of link prediction in general complex network is by Lu et al. \cite{lu2011link} and more specifically for social network we refer to Hasan et al. \cite{al2011survey}. We have generalized topology based methods like Katz (1953) \cite{katz1953new}, proven successful for social networks \cite{liben2007link}, for higher order links. For network alignment based link prediction we refer Flannick et al \cite{flannick2008algorithms} and Xie et al. \cite{xie2012prioritizing}.
\vspace{-0.3em}

The rest of the paper is as follows: Section~\ref{sec:probstatement} describes the problem statement, Section~\ref{sec:methods} describes the topology based methods, in Section~\ref{sec:experiment} the experiments conducted are described and results are discussed, followed by conclusion in section~\ref{sec:conclu}.
\vspace{-0.8em}

\section{Problems and Preliminaries}\label{sec:probstatement}
\subsection{Problem Definition}

In this paper we consider the scenario where we have a set of individuals or social \textit{actors}. These \textit{actors} self assemble themselves into \textit{groups} to perform tasks at hand or gather for an event. A \textit{group} therefore, is a subset of all the \textit{actors}. Membership of a \textit{group} can change over time. An \textit{actor} can leave or join a \textit{group}, resulting in changes in \textit{group} membership. When two \textit{actors} work or gather together in the same \textit{group} they develop social tie. These social ties therefore, become the edges in the social \textit{network of actors} (NOA). Moreover, the \textit{actors} that are the part of multiple groups act as ties between groups resulting in a \textit{network of groups} (NOG) (as we had mentioned in introduction). Given a past history of \textit{groups} formed our problem is to predict \textit{groups} that are likely to form in future by two different evolutionary processes described as follows. Given a \textit{group}, it can absorb an \textit{actor} outside this \textit{group} to form a new group in future. This process is called \textit{incremental accretion} (IA) and the group formed by this process is called \textit{incremental group} (IG) (Figure~\ref{fig:basicprocessesfig}). In the second process, rather than all the members of a given \textit{group}, only a subset of them  absorb an actor outside the group to form another \textit{group}. This process is \textit{subgroup accretion} (SA) and the corresponding group formed is called \textit{subincremental group} (SG) (Figure~\ref{fig:basicprocessesfig}). Note that it is possible that SG and/or IG might have been previously observed or not observed in history. We therefore restrict ourselves to predict only the IG and SG type groups formed by IA and SA processes respectively by assigning prediction scores to them. \vspace{-0.5em}

An example of such a scenario is collaborations among authors to work on publication. As authors write papers they develop social relations with each other. As authors works in multiple research collaborations they become intermediates between these different research collaborations. Related example can be open-source software development teams. 
\vspace{-0.5em}

\subsection{Problem Statement}

We have a set of $n$ \textit{actors} $V = \{v_1,v_2,...,v_n\}$. A subset of these \textit{actors} form a \textit{group}. We have a collection of $m$ such groups observed in past, denoted by $G = \{g_1,g_2,...,g_m\}$ where $g_i \subseteq V $ represents the $i^{th}$ \textit{group}. Cardinality $c_i = |g_i|$ of a group is the number of actors part of it. We have two networks. First, NOG is a hypergraph \cite{berge1973graphs} represented as a set $N_g = (V,G)$ with $G$ as the hyperedges over the vertex set $V$. We also have an incidence matrix $\mathbf{H}$ for $N_g$ of size $(|V| \times |G|)$ with elements defined as $\mathbf{H}(v,g)=1$ if $v \in g$ else $0$. Second, NOA is a graph, $N_a = (V,E)$ where $E=\{e_1,..,e_w\}$ are the dyadic edges defined over vertex set $V$. Adjacency matrix $\mathbf{A}$ of size $(|V| \times |V|)$ for $N_a$ has elements $\mathbf{A}(p,q)=1$ for $(p,q) \in V$ such that $\exists i, \{p,q\} \subseteq g_i$ else $0$.

In IA, a \textit{group} $g_i \in G$ can absorb an actor $a \in \{V-g_i\}$ to produce $g^a_i=\{g_i \cup a \}$. Let $g^{in}_{i} =  \{g^a_i\}_{a \in \{V-g_i\}}$ be the set of all the IGs for $i^{th}$ \textit{group}. Our aim therefore, in IA problem is to predict a score to each of the IGs in set $g^{in}_{i}$ $ \forall i\in\{1,2,..,m\}$. 
\vspace{-1.0em}

Considering the second case of SA problem. We define a proper subgroup of $g_i$ as $s_i \subset g_i$ where $s_i \neq \phi$. A subgroup $s_i$ can absorb an actor $a \in \{V-s_i\}$ to produce $s^a_i=\{s_i \cup a \}$. Let $g^{sa}_{i} = \{s^a_i\}_{a \in \{V-s_i\}}$ be the set of all the SGs for $i^{th}$ \textit{group}. Our aim therefore, in SA problem is to predict a score to each of the SGs in set $g^{sa}_{i}$ $ \forall i\in\{1,2,..,m\}$.
\vspace{-0.8em}

\section{Methods}\label{sec:methods}
In this section we describe three methods to solve the problems described in the previous section. Each method models the affinity of a given \textit{group} towards an \textit{actor} outside the group in different ways. As pointed out earlier, these methods are inspired from the existing dyadic link prediction (DLP) techniques as well as sociology theories. Success of topology based DLP methods encouraged us to focus on topology derived methods. First, method is a generalization of unsupervised path counting based DLP methods to predict the IGs and SGs. The second approach, is a semi-supervised learning based on network alignment algorithms \cite{flannick2008algorithms}. It captures the cycles that pass through both the given group and the rest of graph. The third approach, uses a semi-supervised hypergraph label propagation approach. Each of these methods provide a score $\mathbf{S}(i,j)$ between $i^{th}$ \textit{group} and $j^{th}$ \textit{actor}, representing the similarity or affinity between them.  
\vspace{-0.5em}

\begin{figure*}
\centering
\includegraphics[width=12cm, height=6cm]{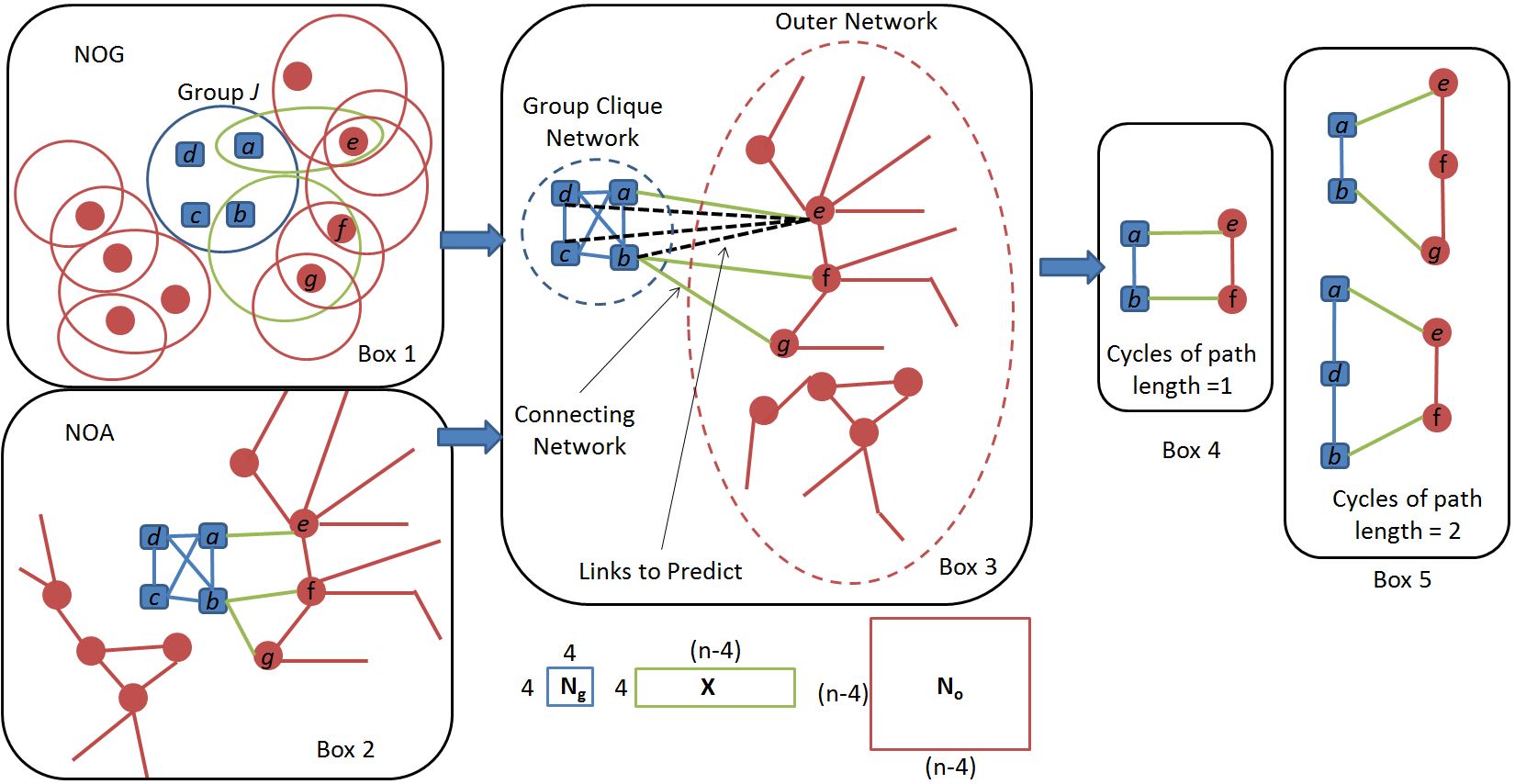}
\caption{Example illustrating different networks used in BRWS for a sample group J consisting of actors $\{a,b,c,d\}$. Box 1 and Box 2 shows the NOG and NOA around Group J actors. In Box 3, we have the 3 network: group clique network (blue), the outer network (red) and bipartite inter network (green). Adjacency matrices used in BRWS are at bottom. Box 4 and Box 5 are example of some cycles of different maximum path length involving actor \textit{e}. These example cycles (along with other cycles not shown) are used in BRWS to calculate scored of the black dotted edges from group members to actor \textit{e}. Note for top cycle in box 5 has path length of 2 over outer network and 1 over clique network. However, these lengths are vice versa for the bottom cycle.}\label{fig:BIRW}
\label{overflow}
\vspace{-0.2em}
\end{figure*}

\subsection{Generalized Katz Score (GKS)}

Among the similarity score based methods in DLP, methods based on counting ensemble of paths have been most successful. More specifically Katz (1953) \cite{katz1953new} measure has been shown to outperform all other similarity scores \cite{liben2007link}. Given the NOA graph $N_a$ the Katz Score (KS) between any two nodes $i$ and $j$ is defined as:
\vspace{-0.4em}
\begin{equation} 
\mathbf{K}(i,j) = \sum_{l=1}^{\infty} \beta^{l}|paths_{i,j}^{(l)}|
\end{equation}
where $\mathbf{K}$ is $(|V|\times|V|)$ size matrix containing KS for different pair of vertices, $|paths_{i,j}^{(l)}|$ is number of $l$ length paths between the nodes $i$ and $j$, and $\beta \in (0,1)$ is parameter that controls the extent to which long paths are penalized. From a sociology perspective number of path between any two nodes capture their social proximity. For example two scientists, in a co-authorship network, who are close to each other in this network, should have many common colleagues or are in similar social circles. Therefore, they are more likely to collaborate. Also smaller length paths reflect greater proximity and are more important, hence, are less penalized or contribute more. In matrix terms it is calculated as: 
\vspace{-0.3em}
\begin{equation} 
\mathbf{K} = \sum_{l=1}^{\infty} \beta^{l}\mathbf{A}^{l}
\end{equation}
where $\mathbf{A}$ is the adjacency matrix for $N_a$. Success of this subtle KS method has encouraged us to generalize it to capture the affinity of a \textit{group} for an \textit{actor} outside it. We therefore, define the Generalized Katz Score (GKS) between $i^{th} group$ and $j^{th} actor$ node as follows:
\begin{equation}
\mathbf{S}(i,j) = \frac{1}{c} \sum_{p \in g_i}\sum_{l=1}^{\infty} \beta^{l}|paths_{p,j}^{(l)}| \\
= \frac{1}{c} \sum_{p \in g_i} \mathbf{K}(p,j)
\end{equation}
where $g_i$ is the $i^{th} group$ of cardinality $c$. This score is the average proximity, measured as KS, of different \textit{actors} within the group to a given external \textit{actor}. For higher order groups the proportion of the group members close to an external individual is more relevant. Therefore, taking average is intuitive, as it tells that on an average how close is this external individual to the given group. It captures the chances he might get absorbed in this group by taking into account the size of group.
\vspace{-0.8em}

\subsection{Bi-random Walk Score (BRWS)}
GKS makes use of the communication paths to measure the affinity of the each group members to a person outside the group separately. In this section, we model the scenario when outside individual is know by multiple members of the group. Consider the group \textit{J} (blue color) in Figure~\ref{fig:BIRW} with four actors \textit{\{a,b,c,d\}} in it. Let us take the  external actor \textit{e} for whom we wish to observe affinity with \textit{J}. We can observe that \textit{e} has a direct link only with the member \textit{a}. But, \textit{b} has a direct link with a neighbor \textit{f} of \textit{e}. Though, there is no direct relation between \textit{b} and \textit{e}, but they have an indirect link through actors both internal (\textit{a}) as well as external (\textit{f}) to the group. To quantify the affinity between \textit{b} and \textit{e}, quantification process should be guided by both these internal and external links. We model this intuition in a holistic way by capturing the cycles like \{\textit{a} $\to$ \textit{b} $\to$ \textit{f} $\to$ \textit{e} $\to$ \textit{a}\}. Instead of simply counting, we use these cycles to learn the affinity of group member to an external actor. We cast this scenario as an alignment problem \cite{flannick2008algorithms} where the nodes within group have to be aligned with external nodes. One of the recently proposed algorithm, by Xie et al. \cite{xie2012prioritizing}, for global network alignment fits very well to our problem with the following modifications.
\vspace{-0.5em}

Again consider the group in Figure~\ref{fig:BIRW}. We take the clique network of the group and the network outside the group and place them apart as shown in box 3 of Figure~\ref{fig:BIRW}. Next we consider three different networks. First network is the group clique network with adjacency matrix $\mathbf{N_{g}}$ of size $(c \times c)$. Second network is the network outside this group with adjacency matrix $\mathbf{N_{o}}$ of size $((n-c) \times (n-c))$. Third network is the inter-network which connects group member nodes to nodes external to group. Its adjacency matrix is $\mathbf{X}$ of size $(c \times (n-c))$. Our aim is to learn the matrix $\mathbf{R}$ whose each entry $\mathbf{R}(p,q)$ contains the affinity score for between the $p^{th}$ group member and the $q^{th}$ actor among the external actor nodes. We define a regularization framework, same as Xie et. al. \cite{xie2012prioritizing}, over the above three networks. The objective function for minimization is:
\vspace{-0.8em}
\begin{multline}
\min_{\mathbf{R}}  \alpha \sum_{u,v,i,j} (\mathbf{N_g} \otimes \mathbf{N_o})_{(i,u),(j,v)} (\mathbf{R}(i,u)-\mathbf{R}(j,v))^2 \\ + (1-\alpha)\sum_{i,u}(\mathbf{R}(i,u)-\mathbf{X}(i,u))^2
\end{multline}
where $\mathbf{N_g} \otimes \mathbf{N_o}$ is the Kronecker product of $\mathbf{N_g}$ and $\mathbf{N_o}$. Each $(\mathbf{N_g} \otimes \mathbf{N_o})_{(i,u),(j,v)}$ is 1 if $\mathbf{N_g}(i,j)=1$ and $\mathbf{N_o}(u,v)=1$, in other words $i^{th}$ and $j^{th}$ group members are linked (which should always be true as group is a clique) and $u^{th}$ and $v^{th}$ external actors are also linked, otherwise 0. The first term in the objective aligns group member $i$ with external actor $u$ and group member $j$ with external actor $v$, if $(i,j)$ are neighbors and $(u,v)$ are also linked. This term therefore, enforces smoothness over $\mathbf{R}$. The second term is a regularization term that uses prior knowledge (like \textit{a} and \textit{e} are already connected in our example) stored in $\mathbf{X}$. $\alpha \in (0,1]$ controls the trade-off between these two competing constraints. As proposed by Xie et. al., the most efficient method to minimize is by the following random-walks based recursive model:
\vspace{-0.5em}
\begin{equation}
\mathbf{R} = \alpha\mathbf{N_g}\mathbf{R}\mathbf{N_o} + (1-\alpha)\mathbf{X}
\end{equation}       
The first term on the right hand side in the $t^{th}$ recursive step becomes $\mathbf{N_g^{t}}\mathbf{R}\mathbf{N_o^{t}}$. This term mimics a random walks across the group network, inter network and the outer network. For $t=1$ it represents cycles with group network path and outer network path length at most 1 as shown in Figure~\ref{fig:BIRW}. In general in $t^{th}$ step it captures cycles with path of length at most $t$ in both clique and outer network. Notice the decay factor $\alpha$ penalizes the larger path length cycles recursively. For further algorithmic details we request to refer to Xie et. al.\cite{xie2012prioritizing}. The exact algorithm used to learn $\mathbf{R}$ for a given group is described in Appendix. Let $\mathbf{R_i}$ represent the learned $\mathbf{R}$ for the group $g_{i}$. Then the affinity of group $g_i$ for $j^{th}$ external actor is:
\vspace{-0.5em}
\begin{equation}
\mathbf{S}(i,j) = \frac{1}{c} \sum_{p=1}^{c} \mathbf{R_i}(p,q)
\end{equation}
where $q$ is the index in outer network $\mathbf{N_o}$ for $i^{th}$ actor in original NOA ($\mathbf{A}$). We therefore, learned  affinity $\mathbf{S(i,j)}$, for $i^{th}$ group and external actor, supervised using the existing connection of the group members to outer actor network. 
\vspace{-0.8em}

\subsection{Group Label Propagation Score (GLPS)}

In the previous sections we developed methods that capture paths and cycles over the NOA but does not takes into account the NOG. In this section we develop label propagation based score which takes into account the hypergraph structure of the NOG. Intuitively, we start by giving some initial labels only to the members of a given $i^{th}$ group $g_i$. These labels then diffuse by random-walks through the hypergraph structure of the NOG. Once the random-walks stabilize, the final label for each external vertex is treated as its affinity score for the given group. The final label at a given external actor vertex represents the chances a random walk originating from group member nodes might end up at this vertex. Therefore, modeling a network guided similarity between the group and the external actor.
\vspace{-0.5em}

To realize the above label diffusion as a hypergraph-based learning task. Let $\mathbf{y}$ be the vector of initial labels to the vertices of the NOG hypergraph $N_g(V,G)$ with incidence matrix $\mathbf{H}$. For a \enquote{given} group $g_i$ we have $\mathbf{y}(v)=1$ if $v \in g_i$ else $\mathbf{y}(v)=0$. We learn the final label vector $\mathbf{f}$. In order to take into account the hypergraph structure we want that the members (vertices) within \enquote{any} group (hyperedge) finally get same labels. Also we want the vertices of \enquote{given} group retain their initial labels. We capture these aims in the following cost minimization objective:
\vspace{-0.5em}
\begin{multline}
\min_{\mathbf{f}} \frac{1}{2}\sum_{g\in G}\sum_{u,v\in g} \frac{\mathbf{w}(g)\mathbf{H}(u,g)\mathbf{H}(v,g)}{\mathbf{\delta}(g)} \left( \frac{\mathbf{f}(u)}{\sqrt{\mathbf{d}(u)}} - \frac{\mathbf{f}(v)}{\sqrt{\mathbf{d}(v)}} \right)^2 \\ + \mu \|\mathbf{f} - \mathbf{y}\|^2
\end{multline}
where, $\mathbf{w}$ is vector whose entries contain hyperedges weights, $\mathbf{d}$ is vector containing vertex degrees such that for as vertex $v$: $\mathbf{d}(v)=\sum_{g\in G|v \in g}\mathbf{w}(g)$ and $\mathbf{\delta}$ is vector containing hyperedge degrees such that for an edge $g$: $\mathbf{\delta}(g)=\sum_{v \in V}\mathbf{H}(v,g)$. The first term is a smoothing term which makes sure that vertices within the same hyperedge have the same scores. So the more number of common hyperedges they are part of the more similar their score becomes. For example if two authors (vertex) have written several papers (hyperedge) together then they are more similar and should be assigned same scores. This term therefore, enforces the hypergraph structure while learning labels. The second term measures the difference between the given labels and the final vertex scores. The parameter $\mu$ then controls the degree of diffusion. In more compact matrix representation:
\begin{equation} \label{eq:opt}
\min_{\mathbf{f}} \mathbf{f}^T \mathbf{L_h} \mathbf{f} + \mu \|\mathbf{f} - \mathbf{y}\|^2
\end{equation}
where, 
\begin{equation}
\mathbf{L_h} = \mathbf{I} - \mathbf{D_v^{-1/2}} \mathbf{H}\mathbf{W}\mathbf{D_e^{-1}}\mathbf{H}^{T}\mathbf{D_v^{-1/2}}
\end{equation} 
is the normalized hypergraph laplacian \cite{zhou2006learning}, where $\mathbf{D_e}$ and $\mathbf{D_v}$ are diagonal matrices consisting of hyperedges and vertex degrees, with ($|G|\times|G|$) and ($|V|\times|V|$) sizes, respectively. $\mathbf{W}$ is the ($|G|\times|G|$) diagonal matrix containing weights of hyperedges. It is easy to show \cite{zhou2006learning} that the solution to equation~\ref{eq:opt} is equivalent to solving the following linear system:
\begin{equation}
\mathbf{f^{*}_i} = (1-\alpha)(\mathbf{I} - \alpha\mathbf{\theta})^{-1}\mathbf{y}, 
\end{equation}
where $\alpha=1/(1+\mu)$, $\mathbf{\theta} = \mathbf{D_v^{-1/2}} \mathbf{H}\mathbf{W}\mathbf{D_e^{-1}}\mathbf{H}^{T}\mathbf{D_v^{-1/2}}$ and $\mathbf{f^{*}_i}$ is the final label vector learned for the $i^{th}$ group. The $j^{th}$ entry in $\mathbf{f^{*}_i}$ quantifies the affinity between of $i^{th}$ group for the $j^{th}$ actor. Therefore, we have:
\begin{equation}
\mathbf{S}(i,j) = \mathbf{f^{*}_i}(j).
\end{equation}

\vspace{-0.8em}  
Since, our aim is to predict a score for each of the IGs $g_i^{in}$ (or SGs $g_i^{sa}$) for all $i\in\{1,....,m\}$. We treat the affinity of group $g_i$ for $j^{th}$ actor ($\mathbf{S}(i,j)$) as the score which reflects the possibility of IG $g^{a}_i \in g_i^{in}$ (actor $a$ is $v_j$) being formed in future. In summary, the prediction score for $g^{a}_i$ (where $a$ is $v_j$) is taken as $\mathbf{S}(i,j)$. In fact we also assign $\mathbf{S}(i,j)$ as the score for SG $s^{a}_i$, same as that for IG $g^{a}_i$. Note that one can build more complicated scores, eg: weighted by the size of subgroup, etc. But for this study we restrict ourselves to this simplified scenario. In experimentation section we shall further discuss how to get the ranked list of most likely future groups using these prediction scores.        
\vspace{-0.8em}  
%

\begin{table} 
\vspace{-0.5em}
\centering
\caption{\textit{Splits} with fixed boundary year}\label{table:split1}
{\renewcommand{\arraystretch}{1.25}%
\begin{tabular}{ |c|c|c|c| }
\hline
Boundary Yr & Split No. & Train & Test \\ \hline
1995 & A.1 & 1992-95 (4 yrs) & 1996-98 (3 yrs) \\ \hline
1995 & A.2 & 1993-95 (3 yrs) & 1996-98 (3 yrs) \\ \hline
1995 & A.3 & 1993-95 (3 yrs) & 1996-99(4 yrs) \\ \hline
2000 & A.4 & 1997-00 (4 yrs) & 2001-03 (3 yrs) \\ \hline
2000 & A.5 & 1998-00 (3 yrs) & 2001-03 (3 yrs) \\ \hline
2000 & A.6 & 1998-00 (3 yrs) & 2001-04 (4 yrs) \\ \hline
2005 & A.7 & 2002-05 (4 yrs) & 2006-08 (3 yrs) \\ \hline
2005 &A.8 & 2003-05 (3 yrs) & 2006-08 (3 yrs) \\ \hline
2005 &A.9 & 2003-05 (3 yrs) & 2006-09 (4 yrs) \\ \hline
2007 & \textbf{Main Split} & 2003-07 (5 yrs) & 2008-10 (3 yrs) \\ \hline
\end{tabular}}
\vspace{-0.5em}
\end{table}

\begin{table*} 
\centering
\caption{Incremental Statistics of Testing Periods for the \textit{Splits} in Table~\ref{table:split1}}\label{table:fix_bdry_incr_stats}
{\renewcommand{\arraystretch}{1.25}%
\begin{tabular}{ |c|c|c|c|c|c|c|c|c|c|c|c|c|c| }
\hline
Split No. & \rot{\parbox{1.6cm}{$\#$ New Actor \\ Groups (NAG)}} & \rot{\parbox{1.6cm}{$\#$ Old Actor \\ Groups (OAG)}} & \rot{$\#$ Total Groups } & \rot{\parbox{1.6cm}{$\%$ New Actor \\ Groups}} & \rot{\parbox{1.6cm}{$\%$ Old Actor \\ Groups }} & \rot{$\#$ Old IGs } & \rot{$\#$ New IGs} & \rot{Total IGs} & \rot{\parbox{1.6cm}{$\%$ (of OAG)\\ New IGs}} & \rot{\parbox{1.6cm}{$\%$ (of OAG)\\ Old IGs}} & \rot{\parbox{1.6cm}{$\%$ (of OAG)\\ Total IGs}} & \rot{\parbox{1.6cm}{$\%$ (of Total IGs) \\ New IGs}} & \rot{\parbox{1.6cm}{$\%$ (of Total IGs) \\ Old IGs}}  \\ \hline
A.1 & 7863 & 2343 & 10206 & 77.043 & 22.95 & 113 & 386 & 499 & 16.47 & 4.82 & 21.29 & 77.36 & 22.64 \\ \hline
A.2 & 8004 & 2202 & 10206 & 78.42 & 21.57 & 81 & 350 & 431 & 15.89 & 3.68 & 19.57 & 81.21 & 18.79  \\ \hline
A.3 & 11665 & 2623 & 14288 & 81.64 & 18.36 & 91 & 416 & 507 & 15.86 & 3.47 & 19.33 & 82.05 & 17.94  \\ \hline
A.4 & 14308 & 3629 & 17937 & 79.77 & 20.23 & 166 & 550 & 716 & 15.16 & 4.57 & 19.73 & 76.81 & 23.18  \\ \hline
A.5 & 14543 & 3394 & 17937 & 81.08 & 18.92 & 139 & 470 & 609 & 13.85 & 4.095 & 17.94 & 77.17 & 22.82  \\ \hline
A.6 & 20331 & 3872 & 24203 & 84.00 & 15.99 & 146 & 543 & 689 & 14.02 & 3.77 & 17.79 & 78.81 & 21.20  \\ \hline
A.7 & 25081 & 6351 & 31432 & 79.79 & 20.20 & 285 & 936 & 1221 & 14.74 & 4.49 & 19.23 & 76.65 & 23.34  \\ \hline
A.8 & 25482 & 5950 & 31342 & 81.30 & 18.98 & 230 & 796 & 1026 & 13.38 & 3.86 & 17.24 & 77.58 & 22.41  \\ \hline
A.9 & 34747 & 6827 & 41474 & 83.78 & 16.46 & 244 & 897 & 1141 & 13.14 & 3.57 & 16.71 & 78.61 & 21.38  \\ \hline
\textbf{Avg.}&        &      &       & 80.76 & \textbf{19.30} &     &      &      & 14.72 & 4.04 & \textbf{18.76} & \textbf{78.47} & 21.52 \\\hline
\hline
\textbf{Main Split (IA)} & 25149 & 7624 & 32773 & 23.26 & 76.74 & 308 & 1085 & 1393 & 14.23 & 4.04 & 18.27&77.89&21.11  \\ \hline 
\textbf{Main Split (SA)} &25149 & 7624 & 32773 & 23.26 & 76.73 & 1503 & 3825 & 5328 & 50.17 & 19.71 & \textbf{69.88} & 71.79 & 28.20  \\ \hline 
\end{tabular}}
\vspace{-1.0em}
\end{table*}

\section{Experimental Analysis}\label{sec:experiment}
In the following the first section describes the dataset statistics. Second section is about evaluation metrics used followed by experiments and analysis in third section.
\vspace{-0.8em}  
%


\subsection{Dataset and Statistics}\label{sec:dataset}
In this paper we have used the popular DBLP dataset (publicly available at \cite{tang2007social}) and extracted all the publication from years 1930-2011 for top 10 venues, as listed in \url{http://academic.research.microsoft.com/}, each for 22 different sub-fields of computer science (total 220 top venues). Here we analyze some interesting statistical properties of this dataset which has motivated this research. Each publication is written by a group of authors and the same group can write multiple publications as well. For both, statistics and experiments we have divided the dataset into various training and test periods (\textit{splits}) as shown in the Tables~\ref{table:split1}. In Table~\ref{table:split1}, each row is a split with a fixed end year of training set (boundary year). Table~\ref{table:fix_bdry_incr_stats} contains the statistics for \textit{(sub)incremental} groups present in \enquote{test} period of different splits. In Table~\ref{table:fix_bdry_incr_stats} we refer an actor or group as \textbf{old} if it has been observed in training else we use \textbf{new}. Note in case a group has written multiple papers in a train or test period it is counted only once. We observe that on an average around $20\%$ of the groups formed in test period contain no new authors (old actor groups (OAG)). Out of these upto approx $20\%$ are \textit{incremental} groups (IGs) formed by IA process. Moreover, around $80\%$ of these IGs are new and never observed in training. Also, we notice that approx $70\%$ of the groups (with no new author (OAG)) are \textit{subincremental} groups (SGs) formed by SA. All these percentages are highlighted in the Table~\ref{table:fix_bdry_incr_stats}. These subtle observations indicate that IA and SA processes are responsible for a large portion of the OAGs formed in future. Therefore, modeling these processes is an important step towards higher order link prediction.  Constrained by space limit we have not shown SG statistics (except for the \textbf{main split} in Table~\ref{table:fix_bdry_incr_stats}). Also the number of actors per split roughly range from $30K$ to $100K$ ($>100K$ in \textbf{main split}) and there are $10K$ to $30K$ groups in \enquote{training} period across various splits. 
\vspace{-0.7em}

\subsection{Evaluation Methodology and Experimental Setup}
In this section we describe two different kinds of metrics, \textbf{global} and \textbf{per-group}. The performance of the proposed approaches is evaluated using the training (2004-07) and testing period (2008-10) of \textbf{main split} in Table~\ref{table:split1}. The statistics of main split are shown at bottom of Table~\ref{table:fix_bdry_incr_stats}. For the groups in the training period, each method: GKS, BRWS and GLPS, was run to output the scores for all IGs ($g_i^{in}$) and all SGs ($g_i^{sa}$) for each group ($i\in\{1,....,m\}$). Now consider the whole set containing all the IGs for all groups. As there might be many repeating IGs we shall only consider unique IGs while taking the maximum score among all the repeating IGs. We sort this unique set of IGs by their scores and get the highest scoring top-$N_{top}$ IGs. We do the exact same thing for SG case and get top-$N_{top}$ SGs. Out of these top-$N_{top}$ groups (IG or SG) the performance over test set (2008-10) is compared using the following metrics:   
\vspace{-0.8em}

\begin{equation}\label{eq:GM1}
\text{Precision@$N_{top}$ (IA)} = \frac{\substack{\text{Number of groups correctly predicted} \\ \text{using IA process from top-}  N_{top} \text{ list}}}{N_{top}}\\
\end{equation}
\begin{equation}
\text{Recall@$N_{top}$ (IA)} = \frac{\substack{\text{Number of collaborations correctly predicted} \\ \text{using IA process from top-}  N_{top} \text{ list}}}{\substack{\text{\# of actual IA generated groups}}}\\
\end{equation}
\begin{equation}
\text{Precision@$N_{top}$ (SA)} = \frac{\substack{\text{Number of groups correctly predicted} \\ \text{using SA process from top-}  N_{top} \text{ list}}}{N_{top}}\\
\end{equation}
\begin{equation}\label{eq:GM4}
\text{Recall@$N_{top}$ (SA)} = \frac{\substack{\text{Number of collaborations correctly predicted} \\ \text{using SA process from top-}  N_{top} \text{ list}}}{\substack{\text{\# of actual SA generated groups}}}\\
\end{equation}

The above \textbf{global} metrics (equations~\ref{eq:GM1} to~\ref{eq:GM4}) capture overall predictions across all groups. But often times we are more interested in understanding the future of a single group and how it will evolve in future. For this case we simply sort IGs $g_i^{in}$ (or SGs $g_i^{sa}$) by their scores in descending order, to get the top-$N_{top}^{g}$ IGs (or SGs) for each $i^{th}$ group. Therefore, for this case we define the following metrics: 
\begin{equation}\label{eq:IA1}
i^{th}\text{GroupPrecision@$N_{top}^{g}$ (IA)} = \frac{\substack{\text{Number of groups correctly} \\ \text{predicted by IA of } i^{th} \text{ group} \\ \text{ from top-}  N_{top}^{g} \text{ list}}}{N_{top}^{g}}\\
\end{equation}
\begin{equation}
i^{th}\text{GroupRecall@$N_{top}^{g}$ (IA)} = \frac{\substack{\text{Number of groups correctly} \\ \text{predicted by IA of } i^{th} \text{ group} \\ \text{ from top-}  N_{top}^{g} \text{ list}}}{\substack{\text{\# of actual IA generated}\\ \text{groups from the }i^{th}\text{ group}}}\\
\end{equation}
for each $i^{th}$ group and take average of these metrics to derive the following average metrics:
\begin{equation}\label{eq:PG1}
\text{AvgPrecision@$N_{top}^{g}$ (IA)} = \frac{\substack{\text{Sum of } i^{th}\text{GroupPrecision@$N_{top}^{g}$ (IA)} \\ \text{for all groups in training set}}}
{\substack{{\text{Total \# of groups in training set}}}}\\
\end{equation}
\begin{equation}\label{eq:PG2}
\text{AvgRecall@$N_{top}^{g}$ (IA)} = \frac{\substack{\text{Sum of }i^{th}\text{GroupRecall@$N_{top}^{g}$ (IA)} \\ \text{for all groups in training set}}}
{\substack{{\text{Total \# of groups in training set}}}}\\
\end{equation}

We refer to the metrics in equations~\ref{eq:PG1} to~\ref{eq:PG2} as the \textbf{per-group} metrics.
Metrics analogous to equation~\ref{eq:IA1} to~\ref{eq:PG2}: $i^{th}$GroupPrecision@$N_{top}^{g}$ (SA), $i^{th}$GroupRecall@$N_{top}^{g}$ (SA), AvgPrecision@$N_{top}^{g}$ (SA) and AvgRecall@$N_{top}^{g}$ (SA), are defined for the SA case as well. 
\vspace{-0.5em}

All the three methods GKS, BRWS and GLPS were implemented in MATLAB. For GKS, BRWS and GLPS the parameters $\beta=\{0.1,0.3,0.5,0.7,0.9\}$, $\alpha=\{0.1,0.2,0.4,0.5,0.6,0.8,0.9\}$, and $\mu=\{0.1,0.3,0.5,0.7,0.9\}$, respectively, were tested. Using 10-fold cross-validation the following best values were observed: $\{\beta=0.5,\alpha=0.6,\mu=0.1\}$ for global metrics and $\{\beta=0.5,\alpha=0.6,\mu=0.5\}$ for per-group metrics. All the hypergraphs and graphs considered in any of the methods are all unweighted and $l\leq4$ is only considered in GKS. Experiments were run using Intel Core i7 (2.8 GHz) CPU with 4 GB RAM.
\vspace{-0.5em}

\subsection{Results and Discussion}
In this section we discuss the results obtained using both, the \textbf{global} and the \textbf{per-group} metrics, for the \textbf{main split}. Note that we omit the results from other splits due to space constraints and moreover, they showed results very similar to the main split. Our \textbf{GKS} method simply extends Katz (1953) \cite{katz1953new} (which is among the most successful DLP methods \cite{liben2007link}). We therefore, consider \textbf{GKS} as a strong baseline for our evaluation. We would also like to mention that we had explored a number of matrix factorization (MF) based DLP methods \cite{al2011survey} like MF, Non-Negative MF, SVD, Tri-MF and their variants with (or without) network regularization and sparsity constraints. But all of them performed trivially in comparison to our methods, so we don't include them. 
\vspace{-0.2em}

We now discuss the results for \textbf{per-group} and \textbf{global} metrics. Notice that in per-group scenario we inspect each group individually and compare its affinity for different actors outside it. Therefore, in this case our comparison is between various \enquote{actors} to ascertain which actors are more likely to join a given group. Whereas in case of global metrics we try to compare scores of different \enquote{groups} (IG or SG) and tell which are more likely to occur in future. Keeping this in mind let us first consider the per-group metrics results in Table~\ref{table:per_group_results} for $N_{top}^{g}=100$. We observe that both \textbf{GLPS} and \textbf{BRWS} outperforms \textbf{GKS} (baseline) consistently for both IA and SA cases. Notice that both \textbf{GLPS} and \textbf{BRWS} are semi-supervised algorithms, with former supervised by the hypergraph structure and the later learns possible cyclic connections from the existing connections between a group with outside actors. In contrast, \textbf{GKS} which simply works on path enumeration based similarity calculation and lacks supervision performs bad. The good performance of \textbf{GLPS} and \textbf{BRWS} suggests that both: (1) hypergraph structure (i.e. how groups as composite entities are connected to each other and therefore, also to the groups in which an external actor participates?) (2) cycles passing through group and an external actor node (i.e. which all group members an external actor knows and through which communication cycles?). Similar, trend is observed in results for global metrics shown in Table~\ref{table:global_results} for $N_{tot}=10000$. Again, \textbf{GLPS} and \textbf{BRWS} perform better than \textbf{GKS}.  However, in this case \textbf{GLPS} does better than \textbf{BRWS} in SA scenario. A possible explanation lies in the fact that \textbf{GLPS} keeps the group as well as subgroup structure intact using the hypergraph model. As an example if we have observed a group P containing actors $\{x,y,z,w\}$ and also its subgroup Q with actors $\{x,y,z\}$. While evaluating group P, \textbf{BRWS} will not consider the existence of the P's subgroup Q as it models the NOA not NOG. Whereas, \textbf{GLPS} models NOG and keep both group as well as subgroup information intact in the hypergraph laplacian of the NOG. This attribute of \textbf{GLPS}, to capture subgroups within other groups, helps it to outperform in SA, where this distinction between groups and its subgroups is quiet critical. 
\vspace{-0.3em}

Another point to mention is the low values for global metrics and precision in case of per group metrics. The reason for this is due to the inherent difficulty of the problem at hand. The number of groups formed by \textit{(sub)incremental} accretion processes (positives occurrence), though a considerable portion of groups in testing period, are much smaller than the total possibilities. Given a group of size $r$ and $n$ as the total number of actors ($>100K$ in main split) in the network. There are $PIA = (n-r)$ and $PSA = \{(2^r-2) \times (n-r)\}$, number of IGs and SGs possible respectively, from the given group. The large number of actors $n(\approx10^5)$ makes $PIA(\approx10^5)$ large and $PSA(\approx10^6)$ (restricting to group size $r_{max} \leq 6$), even much larger, in worst case. Even though, $PIA < PSA$ but the number of IGs formed in test period on an average is much lesser ($20\%$) as compared to ($70\%$) of SGs (~\ref{sec:dataset}). Due to this, in case of per group metric, chances of finding a positive occurrences within $N_{tot}^{g}=100$ groups out of $PIA$ or $PSA$ possibilities, is quiet challenging task. This explains low precision in per group scenario. Global metric scenario is even worse. Assume $m$ (around $30K$ in main split) groups and let us say, for approximation purpose, all are of size $r$. Then there are $GIA = \{m \times PIA\}\approx10^9$ and $GSA=\{m \times PSA$ $\}\approx10^{10}$ number of total IGs and SGs possible (assuming $r_{max} \leq 6$, $m\approx10^4$ and $n\approx10^5$). In case of global metric finding all positive occurrences within just $N_{tot}=10^4$ groups out of the huge $GIA$ and $GSA$ possibilities explains the low global precision scores. (Note above is a rough worst approximation in which we have restricted $r_{max}\leq6$ for illustration. In depth discussion will involve actual group cardinality distribution which we leave due to space limitations.) However, there are a limited number of IA and SA generated groups. We can therefore, hope to cover a significant portion of them within top ranked groups. This makes recall more important measure for us and it attains significantly high values as compared to precision at least for the per group metrics. However, in case of global metrics the number of possibilities are huge, resulting in low values for recall as well.  

\vspace{-1.0em}

\begin{table} 
\centering
\caption{Main Split \textbf{per-group} metrics results}\label{table:per_group_results}
{\renewcommand{\arraystretch}{1.25}%
\begin{tabular}{ |c|c|c|c| }
\hline
& \textbf{GKS} & \textbf{GLPS} & \textbf{BRWS} \\ \hline
AvgPrecision@$100$ (IA) & 0.0210 & 0.0349 & 0.0355\\ \hline
AvgRecall@$100$ (IA) & 0.3176 & 0.6034 & 0.6050\\ \hline
AvgPrecision@$100$ (SA) & 0.0198 & 0.0266 & 0.0271\\ \hline
AvgRecall@$100$ (SA) & 0.2616 & 0.5149 & 0.5135\\ \hline
\end{tabular}}
\vspace{-0.9em}
\end{table}
\begin{table} 
\centering
\caption{Main Split \textbf{global} metrics results}\label{table:global_results}
{\renewcommand{\arraystretch}{1.25}%
\begin{tabular}{ |c|c|c|c| }
\hline
& \textbf{GKS} & \textbf{GLPS} & \textbf{BRWS} \\ \hline
Precision@$10000$ (IA) & 0.0020 & 0.0075 & 0.0134 \\ \hline
Recall@$10000$ (IA) & 0.0144 & 0.0538 & 0.0962\\ \hline
Precision@$10000$ (SA) & 0.0052 & 0.0666 & 0.0327\\ \hline
Recall@$10000$ (SA) & 0.0098 & 0.125 & 0.0614\\ \hline
\end{tabular}}
\vspace{-0.9em}
\end{table}

\section{Conclusions}\label{sec:conclu}
In this work we have addressed the problem of evolution of Small Groups while highlighting its differences with the general problem of community evolution. We found statistically that two \textit{group accretion} processes are behind the formation of a large percentage of future groups given past history of group collaborations. We have built different models that capture these two processes while being motivated from different theories from social science. We treat the problem of future group prediction as a higher-order link prediction task and have developed three topology based methods. Extensive experiments carried out using DBLP dataset show that our methods give good results while predicting future groups.  \\
\\
\textbf{Acknowledgement} This work is supported by NSF Award IIS-1422802 and BBN Contract W911NF-09-2-0053.

\vspace{-0.9em}
\section{Appendix}
\textbf{Bi-random Walk Algorithm} In their paper Xie et. al. \cite{xie2012prioritizing} have provided various versions of Bi-random Walk algorithm. In our work we use the sequential version \textbf{BiRW\_seq}. Input for the algorithm are: the group network adjacency matrix $\mathbf{N_g}$, the inter network matrix $\mathbf{I}$, outer network matrix $\mathbf{N_o}$, the decay parameter $\alpha$, and $\mathit{l_g}$ and $\mathit{l_o}$ are the maximum path length allowed while generating cycles in the group and outer networks respectively. While generating cycles \textbf{BiRW\_seq} does a random-walk on the group network followed by a random-walk on outer network sequentially in each step. In line 3 the algorithm takes a random-walk over the group clique if the length of the path in the group network is still less than $\mathit{l_g}$. A second step is taken on the outer network if path on the outer network covered till now is less than $\mathit{l_o}$ (line 6). In our implementation both $\mathit{l_g}$ and $\mathit{l_o}$ are taken as 4. On reaching  maximum path lengths on both networks $\mathbf{R}$ is returned.
\vspace{-0.8em}

\begin{algorithm}
\caption{\textbf{BiRW\_seq}($\mathbf{N_g}$, $\mathbf{I}$, $\mathbf{N_o}$, $\alpha$, $\mathit{l_g}$,$\mathit{l_o}$)}
\label{alg1}
\begin{algorithmic}[1]
\STATE $\mathbf{R}^{0} = \frac{\mathbf{I}}{sum(\mathbf{I})}$ 
\FORALL{$t=1$ \textbf{to} $\max\{l_g,l_r\}$}
\IF{$t \leq l_g$}
\STATE $\mathbf{R}^{t_{group}} = \alpha\mathbf{N_g}\mathbf{R}^{t-1} + (1-\alpha)\mathbf{I}$
\ENDIF
\IF{$t \leq l_o$}
\STATE $\mathbf{R}^{t} = \alpha\mathbf{R}^{t_{group}}\mathbf{N_o} + (1-\alpha)\mathbf{I}$
\ENDIF
\ENDFOR
\RETURN $(\mathbf{R})$
\vspace{-0.5em}
\end{algorithmic}
\end{algorithm}

\vspace{-1.5em}
\bibliographystyle{IEEEtran}
\bibliography{sigproc}
%
%
%

\end{document}